\documentclass[12pt]{article}
\usepackage{amsfonts,amssymb,epsfig,amsmath}
\newcommand{\Sch}{\textrm{Sch}}

\renewcommand{\text}[1]{#1}

\newcommand{\be}{\begin{equation}}
\newcommand{\ee}{\end{equation}}
\newcommand{\ben}{\begin{displaymath}}
\newcommand{\een}{\end{displaymath}}
\newcommand{\bea}{\begin{eqnarray}}
\newcommand{\eea}{\end{eqnarray}}
\newcommand{\bean}{\begin{eqnarray*}}
\newcommand{\eean}{\end{eqnarray*}}
\newcommand{\nn}{\nonumber \\}
\newcommand{\ba}{\begin{array}}
\newcommand{\ea}{\end{array}}
\newcommand{\bi}{\begin{itemize}}
\newcommand{\ei}{\end{itemize}}

\begin{document}

\makeatletter
\renewcommand{\theequation}{\thesection.\arabic{equation}}
\@addtoreset{equation}{section}
\makeatother

\baselineskip 18pt

\begin{titlepage}

\vfill

\begin{flushright}
KIAS-P09035\\
AEI-2009-052
\end{flushright}

\vfill

\begin{center}
   \baselineskip=16pt
   \begin{Large}\textbf{
        Non-relativistic M-Theory solutions\\*[.3cm] based on K\"ahler-Einstein spaces}
   \end{Large}
   \vskip 1.5cm
    Eoin \'{O} Colg\'{a}in$^1$, Oscar Varela$^2$ and Hossein Yavartanoo$^{1, 3}$
   \vskip .6cm
     \begin{small}
  \textit{ $^1$ Korea Institute for Advanced Study, \\
	Seoul 130-722, Korea}
        \end{small}\\*[.4cm]
       \begin{small}
      \textit{ $^2$ AEI, Max-Planck-Institut f\"ur Gravitationsphysik, \\
	Am M\"uhlenberg 1, D-14476 Potsdam, Germany}
         \end{small}
         \\*[.4cm]
         \begin{small}
          \textit{$^3$ Jefferson Physical Laboratory, \\ Harvard University, Cambridge, MA 02138, USA}
          \end{small}
   \end{center}

\vfill

\begin{center}
\textbf{Abstract}
\end{center}

\begin{quote}
We present new families of non-supersymmetric solutions of $D=11$ supergravity with non-relativistic symmetry,  based on six-dimensional K\"{a}hler-Einstein manifolds. In constructing these solutions, we make use of a consistent reduction to a five-dimensional gravity theory coupled to a massive scalar and vector field. This theory admits a non-relativistic CFT dual with dynamical exponent $z=4$, which may be uplifted to $D=11$ supergravity. Finally, we generalise this solution and find new solutions with various $z$, including $z=2$.
\end{quote}

\vfill

\end{titlepage}

\setcounter{equation}{0}

\section{Introduction}

Over the past year, non-relativistic conformal (NRC) field theories have attracted a lot of attention, primarily driven by the prospect of tailoring the AdS/CFT correspondence so that it may be used as a tool to describe condensed matter systems in a laboratory environment. These systems are described by Schr\"{o}dinger symmetry, which is a non-relativistic version of conformal symmetry. The corresponding algebra is generated by Galilean transformations, an anisotropic scaling of space, $\mathbf{x} \rightarrow \lambda \mathbf{x}$, and time, $x^{+} \rightarrow \lambda^{z} x^{+}$, where $z >0$ is a real number usually referred to as the {\it dynamical exponent}, and an additional special conformal transformation when $z=2$. For NRC field theories with one time and $d$ spatial dimensions, the corresponding symmetry algebra  will be denoted $\Sch_z(1,d)$.

Gravity duals for NRC field theories were initially proposed in \cite{Son:2008ye,Balasubramanian:2008dm} and were subsequently embedded in type IIB in \cite{Adams:2008wt,Herzog:2008wg,Maldacena:2008wh}  and $D=11$ supergravity in \cite{Gauntlett:2009zw}. The IIB solutions of \cite{Adams:2008wt,Herzog:2008wg,Maldacena:2008wh} with $z=2$ are obtained by coordinate transformations which deform the three-form flux, but in the process break supersymmetry.  Other techniques that have been employed in the construction of NRC gravity duals in type IIB and $D=11$ supergravity include metric deformations \cite{Hartnoll:2008rs} and uplift of suitable solutions of the lower dimensional theories to which the $D=10, 11$ supergravities on Sasaki-Einstein manifolds consistently truncate \cite{Maldacena:2008wh,Gauntlett:2009zw}. Some solutions obtained by these two methods do preserve supersymmetry  \cite{Hartnoll:2008rs,Donos:2009en}. Solutions pursued via uplift turn out to permit only set dynamical exponents, whereas more general constructions,
still based on Sasaki-Einstein spaces \cite{Donos:2009en,Bobev:2009mw,Donos:2009xc}, allow for richer classes of solutions with many different values of $z$, including $z=2$ . For a selection of other works on gravity duals of NRC field theories in various dimensions, both supersymmetric and non-supersymmetric, see \cite{otherworks}.

In all these cases, the $D=10$ or $D=11$ metric dual to an NRC field theory in spatial dimension $d$ corresponds to a deformation of a given $D$--dimensional solution containing $(d+3)$-dimensional Anti-de Sitter space, that breaks the original  $AdS_{d+3}$ isometry $so(2,d+2)$ down to its $\Sch_z(1,d)$ subalgebra. The purpose of this paper is to obtain $D=11$ supergravity solutions with $\Sch_z(1,2)$ symmetry, associated to the $AdS_5 \times KE_6$ class of  $D=11$ supergravity solutions with  $KE_6$ a six-dimensional K\"ahler-Einstein space of positive curvature \cite{Dolan:1984hv,Pope:1988xj}. Interestingly enough, despite the lack of supersymmetry of the general $AdS_5 \times KE_6$ solution\footnote{See \cite{Gauntlett:2004zh} for the classification of the superymmetric M-Theory solutions containing $AdS_5$.} for arbitrary $KE_6$, the special case when $KE_6$ is $CP^3$ has recently been shown to be classically stable \cite{Martin:2008pf}. We expect our $\Sch_z(1,2)$--invariant solutions, dual to NRC field theories in spatial dimension $d=2$, to inherit the non-supersymmetric character of the original $AdS_5 \times KE_6$ solutions.

As mentioned earlier, the first examples of gravitational solutions dual to NRC field theories were found in lower-dimensional theories of gravity coupled to a massive vector field \cite{Son:2008ye}. One benefit of much recent work on consistent Kaluza-Klein (KK) truncations \cite{Gauntlett:2007ma,Gauntlett:2006ai,Gauntlett:2007sm} is that these solutions may be uplifted to type IIB \cite{Maldacena:2008wh} and $D=11$ supergravity settings \cite{Gauntlett:2009zw}. In a similar fashion, we will first show, in section \ref{SecConsistent}, that there exists a consistent KK truncation of $D=11$ supergravity on $KE_6$  to a $D=5$ theory involving a massive vector and a massive scalar. We subsequently uplift, in section \ref{NRCuplift}, a solution to the $D=5$ theory  to eleven-dimensions to find a new M-Theory solution with dynamical exponent $z=4$. In section \ref{General} we perform a generalisation to a class of NRC solutions obtained as deformations of the original  $AdS_5 \times KE_6$ solution that, in general,  cannot be obtained from uplift. In this class, we will find new $\Sch_z(1,2)$--invariant M-Theory solutions with different dynamical exponents $z$, including $z=2$. Like the analog constructions in \cite{Hartnoll:2008rs,Donos:2009en,Bobev:2009mw,Donos:2009xc}, the metric of all these solutions will maintain the $KE_6$ part of the original $AdS_5 \times KE_6$. Further generalisations should be possible  allowing for more general internal geometries \cite{progress}.

The $AdS_5 \times KE_6$ geometries that we take as  starting point for our analysis are solutions to the equations of motion of $D=11$ supergravity,
\begin{eqnarray}
&& dG_4 =0 \; , \label{11DBianchi} \\
&& d *_{11} G_4 +\tfrac{1}{2} G_4 \wedge  G_4 =0 \; , \label{11DG4eom} \\
&& R_{AB}= \tfrac{1}{12} G_{A
C_1C_2C_3}G{_{B}}{^{C_1C_2C_3}}-
\tfrac{1}{144}g_{AB} G_{C_1C_2C_3C_4} G^{C_1C_2C_3C_4} =0 \; , \label{11DEinstein}
\end{eqnarray}
with metric and four-form given, respectively, by
\begin{eqnarray} \label{backgroundmetric}
&& ds^2_{11} = ds^2(AdS_5) + ds^2(KE_6), \\
&& \label{background4f}
G_4 = c J \wedge J \; .
\end{eqnarray}
Here, $c$ is a constant, $J$ is the K\"ahler form on $KE_6$, and the metrics  $g_{\mu \nu}$ and $g_{m n}$ for $AdS_5$ and $KE_6$, respectively, are normalised so that their with Ricci tensors are
\be \label{back}
R_{\mu \nu} = - 2 c^2 g_{\mu \nu}, \quad R_{mn} = 2 c^2 g_{mn}.
\ee

{\bf Note.} While we were in the process of completing this paper, \cite{Ooguri:2009cv} appeared which, although supersymmetric in the main, section 5 therein has some overlap with our analysis.

%%%%%%%%%%%%%%%%%%%%%%%%%%%%%%%%%%%%%%%%%%%%%%%%%%%%%%%%%%%%%%%%%%
%%%%%%%%%%%%%%%%%%%%%%%%%%%%%%%%%%%%%%%%%%%%%%%%%%%%%%%%%%%%%%%%%%
%%%%%%%%%%%%%%%%%%%%%%%%%%%%%%%%%%%%%%%%%%%%%%%%%%%%%%%%%%%%%%%%%%

\section{Consistent truncation of $D=11$ supergravity on $KE_6$} \label{SecConsistent}

For every general supersymmetric solution $AdS_n \times_w M_{D-n}$, where $\times_w$ denotes warped product, of a $D$-dimensional supergravity theory, there exists a consistent truncation of the $D$-dimensional theory down to a suitable $n$-dimensional pure, massless gauged supergravity  \cite{Gauntlett:2007ma,Gauntlett:2006ai,Gauntlett:2007sm}. For supersymmetric Freund-Rubin backgrounds, the massive supermultiplet containing the breathing mode of the internal space $M_{D-n}$ can also be retained consistently, together with the supergravity multiplet \cite{Gauntlett:2009zw}. In all these cases, the $G$-structure on $M_{D-n}$ specified by supersymmetry plays a crucial role in constructing the KK ansatz which describes the  embedding of the retained $n$-dimensional fields into the $D$-dimensional ones. In the case at hand here, despite the lack of supersymmetry of the $AdS_5 \times KE_6$ background (\ref{backgroundmetric}), (\ref{background4f}), the K\"ahler form $J$ of $KE_6$ will  still allow us to build a KK ansatz that consistently includes massive modes, along the lines of \cite{Gauntlett:2009zw}.

At any rate, there is an argument about which modes one should expect to be able to keep in the truncation of $D=11$ supergravity on $KE_6$. Consider first the particular case when the internal $KE_6$ is $CP^3$, which has isometry group $SU(4)$, and for which the KK spectrum is explicitly known \cite{Martin:2008pf}. Following \cite{Duff:1985jd}, one should be able to truncate consistently the KK tower of $D=11$ supergravity on $CP^3$ to its $SU(4)$ singlet sector. This contains the massless graviton, one massive real scalar and one massive real vector \cite{Martin:2008pf}, both with mass $12c^2$. Now, it is precisely the singlet character of these modes under the relevant $SU(4)$ symmetry of the particular $KE_6=CP^3$ that makes them expected to be universal for all $KE_6$ spaces. We can thus predict a consistent truncation of $D=11$ supergravity on {\it any} $KE_6$ to a $D=5$ theory with the field content quoted above. In particular, no massless vector that could enter the $D=5$ $N=2$ supergravity multiplet along with the metric should be expected to survive the truncation, so the resulting $D=5$ theory should not correspond to a supergravity\footnote{This is to be constrasted with the analog situation for skew-whiffed Freund-Rubin backgrounds: in spite of  also breaking all supersymmetry, they do allow for a consistent truncation to a supergravity theory \cite{Gauntlett:2009zw}.}.

Without much further ado, consider the following KK ansatz
\begin{eqnarray} \label{KK11DonK6metric}
&& ds^{2}_{11} = ds^{2}_{5} + e^{2U} ds^{2}(KE_6), \\
\label{KK11DonK6}
&& G_4 = H_4 + H_2 \wedge J + c J \wedge J \, ,
\end{eqnarray}
where $U$, $H_4$ and $H_2$ are, respectively, a scalar (the breathing mode of the internal $KE_6$), a four-form and a two-form on the external five-dimensional spacetime, with line element $ds^{2}_{5}$, and $J$ is again the K\"ahler form on  $KE_6$. By choosing the coefficient in the $J \wedge J$ term to be the same constant $c$ that appears in the background flux (\ref{background4f}) we are anticipating that this coefficient cannot be turned into a dynamical $D=5$ field without violating the $D=11$ Bianchi identity for $G_4$. Also, one could have tried to add to the KK ansatz (\ref{KK11DonK6})  terms involving the holomorphic (3,0)-form $\Omega$ defining the complex structure on $KE_6$, but it is unclear how to deal with those terms when plugging the ansatz into the $D=11$ equations of motion.

The KK ansatz (\ref{KK11DonK6metric}), (\ref{KK11DonK6}) reduces to the background solution (\ref{backgroundmetric}), (\ref{background4f}) for $U =H_4=H_2=0$, $ds^2_5 =ds^2 (AdS_5)$. More generally, direct substitution of (\ref{KK11DonK6metric}), (\ref{KK11DonK6}) into
(\ref{11DBianchi})--(\ref{11DEinstein}) shows that the KK ansatz also solves the $D=11$ supergravity field equations provided the $D=5$ fields satisfy
\begin{eqnarray}
\label{BianchiH4}
&& dH_4 =0 \, , \\[6pt]
\label{BianchiH2}
&& dH_2 =0 \, , \\[6pt]
\label{eomH2}
&& d(e^{6U} *H_4) +6cH_2  =0  \, , \\[6pt]
\label{eomH4}
&& d(e^{2U} *H_2) +2cH_4 +H_2 \wedge H_2 =0 \, , \\[6pt]
\label{Ueom5D}
 && d(e^{6U} *dU) +2c^2(e^{-2U} -e^{4U}) \textrm{vol}_5 -\tfrac{1}{6} e^{6U} H_4 \wedge *H_4 =0  \, , \\[6pt]
\label{Ein5D}
 &&  R_{\alpha \beta} \ =  \ -2c^2 e^{-8U}  \eta_{\alpha \beta} + 6 \left(\nabla_\beta \nabla_\alpha U
      + \partial_\alpha U \partial_\beta U \right)
  % \nonumber \\
  % && \qquad
 + \tfrac{3}{2} e^{-4U} \left( H_{\alpha \lambda}
      H_{\beta}{}^{\lambda} -\tfrac{1}{6} \eta_{\alpha \beta}
      H_{\lambda \mu} H^{\lambda \mu} \right)
\nonumber \\
   && \qquad  \qquad
   + \tfrac{1}{12}  \left( H_{\alpha \lambda \mu \nu} H_{\beta}{}^{
        \lambda \mu \nu} -\tfrac{1}{12} \eta_{\alpha \beta} H_{\lambda \mu
        \nu \rho} H^{\lambda \mu \nu \rho} \right) \, .
\end{eqnarray}
All the dependence on the internal $KE_6$ drops out, leaving fully-fledged $D=5$ equations of motion for the $D=5$ fields. This shows the consistency of the truncation.

We can now introduce the Lagrangian of the $D=5$ theory and work out the masses of the various fields. First of all, the Bianchi identities (\ref{BianchiH4}), (\ref{BianchiH2}) for $H_4$ and $H_2$ can be trivially solved by introducing a three-form and a one-form potential such that
\begin{eqnarray}
&& \label{eqB3def} H_4 =dB_3 \, ,\\
&& \label{eqB1def} H_2 =dB_1 .
\end{eqnarray}
The Lagrangian that gives rise to the $D=5$ equations of motion (\ref{eomH2})--(\ref{Ein5D}) upon variation of $B_3$, $B_1$, $U$ and the $D=5$ metric $g_{\mu \nu}$ can then be worked out. It reads
\begin{eqnarray} \label{LagStrFrame}
{\cal L} &=& e^{6U} R \ \textrm{vol}_5 + 30 e^{6U} dU \wedge *dU
-\tfrac12 e^{6U} H_4 \wedge * H_4 -\tfrac32 e^{2U} H_2 \wedge * H_2 \nonumber \\ && + 6c^2 \left( 2e^{4U} - e^{-2U} \right) \textrm{vol}_5 -B_1 \wedge \left( 6 c H_4 + H_2 \wedge H_2 \right) \; ,
\end{eqnarray}
or, in terms of the Einstein frame metric $\bar g_{\mu \nu} = e^{4U}g_{\mu \nu}$,
\begin{eqnarray} \label{lagKE6}
{\cal L}_{\textrm{Einstein}} &=& \bar R \ \bar{\textrm{vol}}_5 -18  dU \wedge \bar * dU
-\tfrac12  e^{12U} H_4 \wedge \bar * H_4 -\tfrac32  H_2 \wedge \bar * H_2 \nonumber \\ && + 6c^2 \left( 2e^{-6U} - e^{-12U} \right) \bar{\textrm{vol}}_5 -B_1 \wedge \left( 6 c H_4 + H_2 \wedge H_2 \right) \; ,
\end{eqnarray}
with barred quantities referring to the Einstein frame metric.

It is useful to dualise $B_3$ into a scalar $B$. In order to do this, define $H_5 =dH_4$ and add the piece
\begin{eqnarray} \label{dualeq1}
{\cal L}^\prime = -BH_5
\end{eqnarray}
to the Lagrangian (\ref{lagKE6}). Integrating out $H_4$ we find that it is now given as
\begin{eqnarray} \label{dualeq2}
H_4 = -e^{-12U} \bar * H_1 \; ,
\end{eqnarray}
where we have found it convenient to define
\begin{eqnarray} \label{dualeq3}
H_1 =  dB - 6c B_1 \; .
\end{eqnarray}
Substituting this back into ${\cal L}_{\textrm{Einstein}} + {\cal L}^\prime$ we find the dual Lagrangian
\begin{eqnarray} \label{LdualK3}
{\cal L}_{\textrm{dual}} &=& \bar R \ \bar{\textrm{vol}}_5 -18  dU \wedge \bar *dU
-\tfrac12 e^{-12U} H_1 \wedge \bar  * H_1 -\tfrac32  H_2 \wedge \bar  * H_2 \nonumber \\ && + 6c^2 \left( 2e^{-6U} - e^{-12U} \right) \bar{\textrm{vol}}_5 -B_1 \wedge H_2 \wedge H_2 \; .
\end{eqnarray}

The masses of the $D=5$ fields can now be computed by expanding the Lagrangian (\ref{LdualK3}) about the $AdS_5$ vacuum, keeping up to quadratic terms. Doing this, for $U$ and $B_1$ we find
\begin{equation} \label{masses}
m^2_U = m^2_{B_1} = 12 c^2 \; ,
\end{equation}
while $B$ (the scalar dual to $B_3$) is just a St\"uckelberg field that can be gauged away to give $B_1$ its mass. As anticipated, the $D=5$ theory obtained upon consistent KK truncation of $D=11$ supergravity on $KE_6$, and described by the Lagrangian (\ref{lagKE6}) or (\ref{LdualK3}), contains the $D=5$ metric, one massive scalar and one massive vector with mass  (\ref{masses}). When $KE_6 =CP^3$, the $SU(4)$--neutrality (table 2 of \cite{Martin:2008pf}) and the masses  (tables 3 and 4 of \cite{Martin:2008pf}) of $U$ and $B_1$ show that these are the modes in the $k=0$ level of the $(k+3)(k+4)c^2$ towers of real scalars and real one-forms, respectively.

We are interested in solutions to the $D=5$ field equations (\ref{BianchiH4})--(\ref{Ein5D}) displaying NRC symmetry. Rather than working with the full theory, we will consider a suitable further truncation. There are three further consistent truncations, apparently no longer explained by a group theory argument as the one above. The first is obtained by setting $H_4=H_2=0$, leaving only the five-dimensional metric and the breathing mode $U$. The second, leading to five-dimensional General Relativity with a cosmological constant, is trivially obtained by insisting on $H_4=H_2=0$ and further setting $U=0$. The third, which is the one we are interested in, will be described in the next section.

\section{NRC solutions from uplift} \label{NRCuplift}

It is consistent with the $D=5$ equations of motion to set $H_4= 6c e^{-6U} *B_1$, where the Hodge dual here refers again to the metric appearing in the Lagrangian (\ref{LagStrFrame}), and $B_1$ is defined in (\ref{eqB1def}) . Rather than a further truncation, this just corresponds to gauging away $B_3$ or, alternatively, the St\"uckelberg scalar $B$, as can be seen from equations (\ref{dualeq2}), (\ref{dualeq3}) . The third possible further truncation referred to above is obtained (having gauged away $B_3$) by further setting $U=0$ (and, thus, $H_4= 6c *B_1$) while restricting $B_1$ to light-like configurations,
\begin{equation} \label{lightlike}
B_1 \wedge * B_1 = 0 \; , \quad  H_2 \wedge H_2 = 0 \; .
\end{equation}
In this case, the equations of motion (\ref{eomH2})--(\ref{Ein5D}) reduce to (\ref{lightlike}) together with
\begin{eqnarray} \label{massivevec1}
&& d*H_2 + 12c^2 *B_1 =0 \; ,  \\
&& \label{massivevec2} R_{\alpha \beta} = -2c^2 \eta_{\alpha \beta} + \tfrac{3}{2}  H_{ \alpha \lambda} H_{ \beta}{}^{\lambda} +18c^2 B_{ \alpha} B_{ \beta} \;
\end{eqnarray}
(with $H_2 =dB_1$). Indeed, setting $U=0$ and $H_4= 6c *B_1$, equation (\ref{eomH2}) is identically satisfied; equations  (\ref{eomH4}) and (\ref{Ueom5D}) reduce, respectively, to the second and first conditions in (\ref{lightlike}); equation (\ref{BianchiH4}) is obtained by differentiating (\ref{massivevec1}); and, finally, the Einstein equation (\ref{Ein5D}) reduces to (\ref{massivevec2}).

The equations of motion (\ref{massivevec1}), (\ref{massivevec2}) can be derived from the Lagrangian\footnote{This $D=5$ theory, with even the same mass for the vector $B_1$ if we choose $c= \sqrt{2}$, was first discussed in section 4.2 of \cite{Maldacena:2008wh}, but the $D=5$ parent theories with Lagrangian (\ref{LdualK3}) above and (4.21) of \cite{Maldacena:2008wh} are very different. As in \cite{Maldacena:2008wh, Gauntlett:2009zw}, the Lagrangian (\ref{lagKE6mass}) only reproduces the equations (\ref{massivevec1}), (\ref{massivevec2}) and not the light-like condition (\ref{lightlike}). Since (\ref{lightlike}),  (\ref{massivevec1}), (\ref{massivevec2}) can be consistently obtained upon truncation of $D=11$ supergravity on $KE_6$, any choice of five-dimensional metric and lightlike $B_1$ (thus subject to (\ref{lightlike})) which also solves the equations of motion (\ref{massivevec1}), (\ref{massivevec2}) that derive from the Lagrangian (\ref{lagKE6mass}), can be safely uplifted to $D=11$.}
\begin{eqnarray} \label{lagKE6mass}
{\cal L} = R \ \textrm{vol}_5 +6c^2 \textrm{vol}_5 -\tfrac32  H_2 \wedge * H_2 -18c^ 2 B_1 \wedge *B_1 \; ,
\end{eqnarray}
which was argued in \cite{Son:2008ye} to allow for solutions with metric displaying Schr\"odinger symmetry. These solutions should be supported by a light-like massive vector of the form $B_1 \propto r^z dx^+$ (see \cite{Maldacena:2008wh}), where $z$ is the dynamical exponent, thus immediately satisfying (\ref{lightlike}). Specifically, we look for solutions to (\ref{lightlike}), (\ref{massivevec1}), (\ref{massivevec2}) of the form
\begin{eqnarray} \label{ansatzNRC5d}
&& ds^2_5 = -\alpha^2 r^{2z} (dx^+)^2 +\frac{2}{c^2 r^2} dr^2 + \frac{2}{c^2} r^2 \left( -dx^+dx^- +dx_1^2 +dx_2^2 \right)  \ , \nonumber \\
&& B_1 = \beta r^z dx^+.
\end{eqnarray}
where $\alpha$, $\beta$ and the dynamical exponent $z$ are constants to be determined. The configuration (\ref{ansatzNRC5d}) does satisfy the conditions (\ref{lightlike}) and turns out to also solve the equations (\ref{massivevec1}), (\ref{massivevec2}) provided that
\bea
z(z+2) &=& 24 \; ,  \label{intz} \\
\alpha^2 (z^2-1) &=& \beta^2 (\tfrac{3}{4} z^2 + 18) \label{z=1}.
\eea
Thus, as in \cite{Maldacena:2008wh}, we indeed find solutions for $z=4$ (and $\beta = \frac{\alpha}{\sqrt{2}}$) and $z=-6$ (and $\beta = \frac{\alpha \sqrt{7}}{3}$). By convention $z>0$, so we ignore the latter possibility.

The $z=4$ solution can now be uplifted to $D=11$ with the help of the KK ansatz (\ref{KK11DonK6metric}), (\ref{KK11DonK6}). We find
\begin{eqnarray} \label{11Dsol}
&& ds^2_{11} = -\alpha^2 r^{8} (dx^+)^2 +\frac{2}{c^2} \frac{dr^2}{r^2} + \frac{2}{c^2} r^2 \left( -dx^+dx^- +dx_1^2 +dx_2^2 \right) + ds^ 2(KE_6) \ , \nonumber \\[8pt]
&& G_4 = 12 \tfrac{\alpha}{c^2} r^5 dx^+\wedge dr \wedge dx_1  \wedge dx_2 -2 \sqrt{2}\alpha r^3 dx^+\wedge dr\wedge J +cJ\wedge J \; .
\end{eqnarray}
This is a new (non-supersymmetric) M-Theory solution dual to a NRC field theory in spatial dimension $d=2$ with dynamical exponent $z=4$.
One can generalise this solution and consider  more general ansatze for $D=11$ supergravity solutions dual to $d=2$ non-relativistic conformal field theories with dynamical exponent $z$, where the internal directions still correspond to a $KE_6$ space. We now turn to this point.

%%%%%%%%%%%%%%%%%%%%%%%%%%%%%%%%%%%%%%%%%%%%%%%%%%%%%%%%%%%%%%%%%%%%%%%%%%
%%%%%%%%%%%%%%%%%%%%%%%%%%%%%%%%%%%%%%%%%%%%%%%%%%%%%%%%%%%%%%%%%%%%%%%%%%
%%%%%%%%%%%%%%%%%%%%%%%%%%%%%%%%%%%%%%%%%%%%%%%%%%%%%%%%%%%%%%%%%%%%%%%%%%

\section{Some generalisations} \label{General}

As we have just mentioned, the $D=11$ solution (\ref{11Dsol}) is locally invariant under $\Sch_4 (1,2)$. In particular, the scale invariance acts on coordinates as \cite{Balasubramanian:2008dm}
\be \label{scaletrans} (x^{+},x^{-}, x_i,r) \rightarrow (\lambda^z x^+,\lambda^{2-z} x^{-},\lambda x_i, \lambda^{-1} r) \; , \quad  \; i=1,2  \ee
(with $z=4$ in (\ref{11Dsol})), while leaving the $KE_6$ coordinates unchanged. Following \cite{Hartnoll:2008rs,Donos:2009en}, we can generalise the metric in (\ref{11Dsol}) as:
\begin{eqnarray} \label{ansatzNRC}
ds^2_{11} &=& \frac{2}{c^2} \Big[ - f_0 r^{2z} (dx^+)^2 - r^2 dx^+ (dx^- +r^{z-2} C_1) +\frac{1}{r^2} dr^2  \nonumber \\ && \qquad  +  r^2 \big( dx_1^2 +dx_2^2 \big)  \Big]  + ds^2(KE_6) \ ,
\end{eqnarray}
where $C_1$ is a one-form and $f_0$ a function, both of them defined on the internal $KE_6$. Both $C_1$ and $r^{2z} f_0$, serve the same role of breaking the $SO(2,4)$ isometry of the original $AdS_5 \times KE_6$ metric (\ref{backgroundmetric}) down to $\Sch_z (1,2)$.

An ansatz for the accompanying four-form flux may be constructed by considering the forms invariant under $Sch_z(1,2)$ symmetry (see \cite{Colgain:2009wm}), though the equations of motion constrain the candidate forms.  The specific ansatz we then consider for the four-form flux is
\begin{eqnarray} \label{fluxNRC2}
G_4 &=& -\tfrac{1}{z+2} d( \mu_0 r^{z+2} dx^+ \wedge dx^1\wedge dx^2) -\tfrac{1}{z} d( \mu_2 \wedge r^{z} dx^{+}) +c J \wedge J \;,
\end{eqnarray}
where, in general, $\mu_0$ is a function and $\mu_2$ a two-form, both defined on $KE_6$. The latter can be taken to be proportional to the K\"ahler form on $KE_6$, as for the uplifted $z=4$ solution (\ref{11Dsol}),  but other choices are also possible (see subsection \ref{classSqrt3} below). Indeed, the solution (\ref{11Dsol}) is recovered from (\ref{ansatzNRC}), (\ref{fluxNRC2}) by setting $C_1=0$, $f_0=\tfrac12 c^2 \alpha^2$, $\mu_0 =\tfrac{12 \alpha}{c^2}$ and $\mu_2 =-2\sqrt{2} \alpha J$, for some constant $\alpha$. More generally, the non-trivial mixing of external and $KE_6$ coordinates in the metric (\ref{ansatzNRC}) will prevent it from being obtainable as the uplift of any $D=5$ metric. The requirement that (\ref{ansatzNRC}), (\ref{fluxNRC2}) do solve the equations of motion (\ref{11DBianchi})--(\ref{11DEinstein}) of  $D=11$ supergravity leads to restrictions and relations for $f_0$, $C_1$, $\mu_0$ and $\mu_2$. In the following, we will spell out several interesting cases.

\subsection{A solution with $z=2$}

We can find a $D=11$ supergravity solution with dynamical exponent $z=2$ by setting, for some constant $\alpha$, $f_0 = \frac{ 13 \alpha }{4 c^4}$, choosing  $C_1$ such that $dC_1= \alpha J$, while writing  $\mu_0=\tfrac{12\alpha \sqrt{2}}{c^5}$, $\mu_2= - \tfrac{2 \alpha}{c^3}$ so that the flux (\ref{fluxNRC2}) reads
\begin{eqnarray}
G_4 = \tfrac{12\alpha \sqrt{2}}{c^5} r^{3} dx^+\wedge dr \wedge dx_1  \wedge dx_2 - \tfrac{2 \alpha}{c^3} r dx^+\wedge dr\wedge J +cJ\wedge J \;.
\end{eqnarray}
A generalisation of this solution appeared previously in \cite{Ooguri:2009cv}, where the internal space is a variant of $CP^3$ \cite{Pope:1988xj}.

\subsection{A class of solutions with $z \geq \sqrt{3}$} \label{classSqrt3}

Setting $C_1=0$ in the metric (\ref{ansatzNRC}) and $\mu_0=0$, $\mu_2=0$ in (\ref{fluxNRC2}) (which takes the flux back to its background value (\ref{background4f})), some calculation reveals that the resulting combination of metric and four-form provides a solution of $D=11$ supergravity  if $f_0$ is an eigenfunction of the Laplacian $\Delta_{KE} \equiv *d*d + d * d*$ on $KE_6$ with eigenvalue $ 2(z^2-1) c^2$:
\begin{equation} \label{laplacianonf}
\Delta_{KE} f_0 = 2(z^2-1) c^2 f_0.
\end{equation}
This class of solutions thus provides a $D=11$ counterpart of the Type IIB solutions first discussed in \cite{Hartnoll:2008rs}.

For the particular case $KE_6 =CP^3$, these eigenvalues are   $k(k+3)c^2$, $k=0, 1 , \ldots$, with the corresponding eigenfunctions transforming in the $(k0k)$ irrep of $SU(4)$ \cite{taniguchi,Martin:2008pf}. Ruling out $k=0$, which just corresponds to a space locally isometric to  $AdS_5 \times KE_6$, we have a sequence of families of solutions with dynamical exponents
\begin{equation} \label{zk}
z_k = \sqrt{1+ \tfrac12 k(k+3)} \; , \quad k= 1, 2 \ldots \; ,
\end{equation}
thus obeying the bound
\begin{equation}
z_k \geq \sqrt{3} \; .
\end{equation}
For each $k=1,2,3 \ldots$, this class contains a family of $\textrm{dim}(k0k) = 15, \ 84, \ 300, \ldots$ supergravity solutions with the dynamical exponent $z_k$ in (\ref{zk}).

As noted in \cite{Hartnoll:2008rs}, this class of solutions should be unstable. Stability could be restored in  \cite{Hartnoll:2008rs} by appropriately turning on fluxes. We can try to do the same here by setting, for simplicity, $\mu_2$ to be proportional to the K\"{a}hler form $J$. In this case, only for $z=4$ do we find a solution with metric (\ref{ansatzNRC}) (with $C_1=0$), supported by the flux
\begin{eqnarray}
G_4 = \alpha r^5 dx^+\wedge dr \wedge dx_1  \wedge dx_2 - \tfrac{\alpha c^2}{3\sqrt{2}} r^3 dx^+\wedge dr\wedge J +cJ\wedge J \; ,
\end{eqnarray}
for any constant $\alpha$. In this case, $f_0$ gets shifted by a positive term proportional to $\alpha^2$, which can be tuned to render the solution stable \cite{Hartnoll:2008rs}.  The shifted $f_0$ still fulfils (\ref{laplacianonf}), now with eigenvalue $30c^2$, corresponding to $z=4$. We are unaware, however, of any $KE_6$ space for which this eigenvalue is permissible.

Alternatively, following \cite{Donos:2009en,Bobev:2009mw,Donos:2009xc}, rather than setting $\mu_2$ to be  proportional to the K\"{a}hler form, one may take it to be primitive and transverse\footnote{A $(p,q)$--form $Y^{p,q}$ on a K\"ahler space is said to be primitive if its contraction with the K\"ahler form vanishes, $J^{mn} Y^{p,q}_{mn...} = 0$, and transverse if $* d * Y^{p,q} = 0$.}. Setting, for convenience, $\mu_0 = C_1 = 0$, a calculation shows that the configuration (\ref{ansatzNRC}), (\ref{fluxNRC2}) is a solution to $D=11$ supergravity provided
\bea \label{eqmu2}
\Delta_{KE} f_0 + 2 (z^2-1) c^2 f_0&=& \frac{c^4}{4} |\mu_2|^2 + \frac{c^2}{2 z^2} |d \mu_2|^2, \nn
\Delta_{KE} \mu_2 &=& \tfrac{1}{2} z(z +2)c^2 \mu_2,
\eea
where $|\mu_2|^2 = \tfrac{1}{2!} \mu_{2\;ab} \mu_2^{ab}$, etc.  Now, $f_0$ has devolved the Laplacian eigenvector character upon $\mu_2$, which corresponds to  a two-form eigenfunction with eigenvalue $\tfrac{1}{2} z(z +2)c^2$. In the special case $KE_6=CP^3$, the eigenvalues of the Laplacian acting on transverse, primitive $(1,1)$--forms (respectively, $(2,0)$--forms) are $(k+2)(k+3)c^2$ (respectively, $(k+3)(k+4)c^2$), for $k=0,1, \ldots$ \cite{taniguchi,Martin:2008pf}. We thus see that solutions to (\ref{eqmu2}) correspond to NRC gravity duals with dynamical exponents bounded below by $z \geq -1+ \sqrt{13}$ (respectively, $z \geq 4$), if $\mu_2$ is a chosen to be (the real part of) a $(1,1)$--form (respectively, $(2,0)$--form). See \cite{Donos:2009xc} for a discussion of a solving technique for systems of equations like (\ref{eqmu2}). It would be interesting to study the stability of this class of solutions.

\section{Final comments}

We have constructed solutions of $D=11$ supergravity dual to NRC field theories in 2 spatial dimensions and with different values of the dynamical exponent $z$. They correspond to suitable deformations of the class of solutions $AdS_5 \times KE_6$, that break the $SO(2,4)$ symmetry down to its Schr\"odinger subalgebra $\Sch_z(1,2)$. Important insight was obtained by first dealing with a simpler, particular solution with $z=4$. Specifically, $D=11$ supergravity reduced on the internal $KE_6$ truncates consistently to a $D=5$ gravity theory involving a massive vector. A suitable solution of this theory, with $z=4$, was found and subsequently uplifted to eleven-dimensions. We also discussed a more general class of $D=11$ supergravity solutions, locally invariant under $\Sch_z(1,2)$, that contains this solution, along with  other examples that can no longer be obtained upon uplift. We are able to find explicitly a solution with $z=2$, a class of solutions with dynamical exponents $z \geq \sqrt{3}$, and implicitly, solutions with $z \geq -1+ \sqrt{13}$ and $z \geq 4$.

The Schr\"odinger algebra $\Sch_z(1,d)$ is not the only NRC symmetry one may consider. In fact, there also exists a conformal version of the Galilean algebra that, unlike $\Sch_z(1,d)$, can be obtained as an In\"on\"u-Wigner contraction of the relativistic conformal algebra $so(2,d+2)$. Some issues regarding the Galilean conformal algebra have been recently discussed, including its supersymmetrisation  \cite{Sakaguchi:2009de,deAzcarraga:2009ch,Bagchi:2009ke} and its implementation, both in the dual field theories and the gravity bulk \cite{Bagchi:2009my,Martelli:2009uc}. As pointed out in \cite{Martelli:2009uc}, a drawback of backgrounds with this conformal Galilean symmetry is that, in contrast to $\Sch_z(1,d)$--invariant ones, their metrics exhibit a non-Lorentzian signature. While this would require better understanding, progress on the way NRC symmetries are implemented in gravity duals may be achieved by a systematic characterisation \cite{progress} of Type IIB and M-Theory backgrounds with $\Sch_z(1,d)$ symmetry, for generic values of $z$ and $d$.

\section*{Acknowledgements}

We would like to thank Ido Adam, Dumitru Astefanesei, Jos\'e A. de Azc\'arraga, Jerome Gauntlett, Sean Hartnoll, Hironobu Kihara, Dario Martelli, Carlos N\'u\~nez, Ioannis Papadimitriou, Mar\'\i a J. Rodr\'\i guez and Stefan Theisen for helpful discussions. OV is supported by an Alexander von Humboldt postdoctoral fellowship and, partially, through the Spanish Government research grant FIS2008-01980.


\begin{thebibliography}{99}


%\cite{Son:2008ye}
\bibitem{Son:2008ye}
  D.~T.~Son,
  %``Toward an AdS/cold atoms correspondence: a geometric realization of the
  %Schroedinger symmetry,''
  Phys.\ Rev.\  D {\bf 78} (2008) 046003
  [arXiv:0804.3972 [hep-th]].
  %%CITATION = PHRVA,D78,046003;%%

%\cite{Balasubramanian:2008dm}
\bibitem{Balasubramanian:2008dm}
  K.~Balasubramanian and J.~McGreevy,
  %``Gravity duals for non-relativistic CFTs,''
  Phys.\ Rev.\ Lett.\  {\bf 101}, 061601 (2008)
  [arXiv:0804.4053 [hep-th]].
  %%CITATION = PRLTA,101,061601;%%

%\cite{Adams:2008wt}
\bibitem{Adams:2008wt}
  A.~Adams, K.~Balasubramanian and J.~McGreevy,
  %``Hot Spacetimes for Cold Atoms,''
  JHEP {\bf 0811}, 059 (2008)
  [arXiv:0807.1111 [hep-th]].
  %%CITATION = JHEPA,0811,059;%%


%\cite{Herzog:2008wg}
\bibitem{Herzog:2008wg}
  C.~P.~Herzog, M.~Rangamani and S.~F.~Ross,
  %``Heating up Galilean holography,''
  JHEP {\bf 0811}, 080 (2008)
  [arXiv:0807.1099 [hep-th]].
  %%CITATION = JHEPA,0811,080;%%

%\cite{Maldacena:2008wh}
\bibitem{Maldacena:2008wh}
  J.~Maldacena, D.~Martelli and Y.~Tachikawa,
  %``Comments on string theory backgrounds with non-relativistic conformal
  %symmetry,''
  JHEP {\bf 0810}, 072 (2008)
  [arXiv:0807.1100 [hep-th]].
  %%CITATION = JHEPA,0810,072;%%


%\cite{Gauntlett:2009zw}
\bibitem{Gauntlett:2009zw}
  J.~P.~Gauntlett, S.~Kim, O.~Varela and D.~Waldram,
  %``Consistent supersymmetric Kaluza--Klein truncations with massive modes,''
  arXiv:0901.0676 [hep-th].
  %%CITATION = ARXIV:0901.0676;%%


%\cite{Hartnoll:2008rs}
\bibitem{Hartnoll:2008rs}
  S.~A.~Hartnoll and K.~Yoshida,
  %``Families of IIB duals for nonrelativistic CFTs,''
  JHEP {\bf 0812}, 071 (2008)
  [arXiv:0810.0298 [hep-th]].
  %%CITATION = JHEPA,0812,071;%%


%\cite{Donos:2009en}
\bibitem{Donos:2009en}
  A.~Donos and J.~P.~Gauntlett,
  %``Supersymmetric solutions for non-relativistic holography,''
  JHEP {\bf 0903}, 138 (2009)
  [arXiv:0901.0818 [hep-th]].
  %%CITATION = JHEPA,0903,138;%%

%\cite{Bobev:2009mw}
\bibitem{Bobev:2009mw}
  N.~Bobev, A.~Kundu and K.~Pilch,
  %``Supersymmetric IIB Solutions with Schr\'{o}dinger Symmetry,''
  arXiv:0905.0673 [hep-th].
  %%CITATION = ARXIV:0905.0673;%%

%\cite{Donos:2009xc}
\bibitem{Donos:2009xc}
  A.~Donos and J.~P.~Gauntlett,
  %``Solutions of type IIB and D=11 supergravity with Schrodinger(z) symmetry,''
  arXiv:0905.1098 [hep-th].
  %%CITATION = ARXIV:0905.1098;%%


%\cite{Bobev:2009zf}
\bibitem{otherworks}
  S.~Kachru, X.~Liu and M.~Mulligan,
  %``Gravity Duals of Lifshitz-like Fixed Points,''
  Phys.\ Rev.\  D {\bf 78}, 106005 (2008)
  [arXiv:0808.1725 [hep-th]];
  %%CITATION = PHRVA,D78,106005;%%
  S.~Sekhar Pal,
  %``Towards Gravity solutions of AdS/CMT,''
  arXiv:0808.3232 [hep-th];
  %%CITATION = ARXIV:0808.3232;%%
  C.~Duval, M.~Hassaine and P.~A.~Horvathy,
  %``The geometry of Schr\'odinger symmetry in gravity
  %background/non-relativistic CFT,''
  Annals Phys.\  {\bf 324}, 1158 (2009)
  [arXiv:0809.3128 [hep-th]];
  %%CITATION = APNYA,324,1158;%%
  M.~Schvellinger,
  %``Kerr-AdS black holes and non-relativistic conformal QM theories in diverse
  %dimensions,''
  JHEP {\bf 0812}, 004 (2008)
  [arXiv:0810.3011 [hep-th]];
  %%CITATION = JHEPA,0812,004;%%
  L.~Mazzucato, Y.~Oz and S.~Theisen,
  %``Non-relativistic Branes,''
  JHEP {\bf 0904}, 073 (2009)
  [arXiv:0810.3673 [hep-th]];
  %%CITATION = JHEPA,0904,073;%%
  A.~Adams, A.~Maloney, A.~Sinha and S.~E.~Vazquez,
  %``1/N Effects in Non-Relativistic Gauge-Gravity Duality,''
  JHEP {\bf 0903}, 097 (2009)
  [arXiv:0812.0166 [hep-th]];
  %%CITATION = JHEPA,0903,097;%%
  M.~Taylor,
  %``Non-relativistic holography,''
  arXiv:0812.0530 [hep-th];
  %%CITATION = ARXIV:0812.0530;%%
  S.~S.~Pal,
  %``Anisotropic gravity solutions in AdS/CMT,''
  arXiv:0901.0599 [hep-th];
    %%CITATION = ARXIV:0901.0599;%%
    %
    M.~Alishahiha, A.~Davody and A.~Vahedi,
  %``On AdS/CFT of Galilean Conformal Field Theories,''
  arXiv:0903.3953 [hep-th];
  %%CITATION = ARXIV:0903.3953;%%
  N.~Bobev and A.~Kundu,
  %``Deformations of Holographic Duals to Non-Relativistic CFTs,''
  arXiv:0904.2873 [hep-th];
  S.~S.~Pal,
  %``Non-relativistic supersymmetric Dp branes,''
  arXiv:0904.3620 [hep-th].
  %%CITATION = ARXIV:0904.3620;%%

  %%CITATION = ARXIV:0904.2873;%%


%\cite{Dolan:1984hv}
\bibitem{Dolan:1984hv}
  B.~Dolan,
  %``A New Solution Of D = 11 Supergravity With Internal Isometry Group SU(3) X
  %SU(2) X U(1),''
  Phys.\ Lett.\  B {\bf 140} (1984) 304.
  %%CITATION = PHLTA,B140,304;%%

%\cite{Pope:1988xj}
\bibitem{Pope:1988xj}
  C.~N.~Pope and P.~van Nieuwenhuizen,
  %``COMPACTIFICATIONS OF d = 11 SUPERGRAVITY ON KAHLER MANIFOLDS,''
  Commun.\ Math.\ Phys.\  {\bf 122} (1989) 281.
  %%CITATION = CMPHA,122,281;%%



%\cite{Gauntlett:2004zh}
\bibitem{Gauntlett:2004zh}
  J.~P.~Gauntlett, D.~Martelli, J.~Sparks and D.~Waldram,
  %``Supersymmetric AdS(5) solutions of M-theory,''
  Class.\ Quant.\ Grav.\  {\bf 21} (2004) 4335
  [arXiv:hep-th/0402153].
  %%CITATION = CQGRD,21,4335;%%



%\cite{Martin:2008pf}
\bibitem{Martin:2008pf}
  J.~E.~Martin and H.~S.~Reall,
  %``On the stability and spectrum of non-supersymmetric AdS(5) solutions of
  %M-theory compactified on Kahler-Einstein spaces,''
  JHEP {\bf 0903}, 002 (2009)
  [arXiv:0810.2707 [hep-th]].
  %%CITATION = JHEPA,0903,002;%%



%\cite{Gauntlett:2007ma}
\bibitem{Gauntlett:2007ma}
  J.~P.~Gauntlett and O.~Varela,
  %``Consistent Kaluza-Klein Reductions for General Supersymmetric AdS
  %Solutions,''
  Phys.\ Rev.\  D {\bf 76} (2007) 126007
  [arXiv:0707.2315 [hep-th]].
  %%CITATION = PHRVA,D76,126007;%%


%\cite{Gauntlett:2006ai}
\bibitem{Gauntlett:2006ai}
  J.~P.~Gauntlett, E.~O Colgain and O.~Varela,
  %``Properties of some conformal field theories with M-theory duals,''
  JHEP {\bf 0702}, 049 (2007)
  [arXiv:hep-th/0611219].
  %%CITATION = JHEPA,0702,049;%%

%\cite{Gauntlett:2007sm}
\bibitem{Gauntlett:2007sm}
  J.~P.~Gauntlett and O.~Varela,
  %``$D=5$, $SU(2)\times U(1)$ Gauged Supergravity from $D=11$ Supergravity,''
  JHEP {\bf 0802} (2008) 083
  [arXiv:0712.3560 [hep-th]].
  %%CITATION = JHEPA,0802,083;%%

\bibitem{progress} Work in progress.

%\cite{Ooguri:2009cv}
\bibitem{Ooguri:2009cv}
  H.~Ooguri and C.~S.~Park,
  %``Supersymmetric non-relativistic geometries in M-theory,''
  arXiv:0905.1954 [hep-th].
  %%CITATION = ARXIV:0905.1954;%%



  %\cite{Duff:1985jd}
\bibitem{Duff:1985jd}
  M.~J.~Duff and C.~N.~Pope,
  %``Consistent Truncations In Kaluza--Klein Theories,''
  Nucl.\ Phys.\  B {\bf 255} (1985) 355.
  %%CITATION = NUPHA,B255,355;%%


%\cite{Colgain:2009wm}
\bibitem{Colgain:2009wm}
  E.~O.~Colgain and H.~Yavartanoo,
  %``NR $CFT_3$ duals in M-theory,''
  arXiv:0904.0588 [hep-th].
  %%CITATION = ARXIV:0904.0588;%%


\bibitem{taniguchi}
A. Ikeda and Y. Taniguchi. %``Spectra end eigenforms of the Laplacian on $S^n$ and $P^n(C)$'',
Osaka J. Math \textbf{15} 515 (1978).


%\cite{Sakaguchi:2009de}
\bibitem{Sakaguchi:2009de}
  M.~Sakaguchi,
  %``Super Galilean conformal algebra in AdS/CFT,''
  arXiv:0905.0188 [hep-th].
  %%CITATION = ARXIV:0905.0188;%%



%\cite{deAzcarraga:2009ch}
\bibitem{deAzcarraga:2009ch}
  J.~A.~de Azcarraga and J.~Lukierski,
  %``Galilean Superconformal Symmetries,''
  arXiv:0905.0141 [math-ph].
  %%CITATION = ARXIV:0905.0141;%%

%\cite{Bagchi:2009ke}
\bibitem{Bagchi:2009ke}
  A.~Bagchi and I.~Mandal,
  %``Supersymmetric Extension of Galilean Conformal Algebras,''
  arXiv:0905.0580 [hep-th].
  %%CITATION = ARXIV:0905.0580;%%


%\cite{Bagchi:2009my}
\bibitem{Bagchi:2009my}
  A.~Bagchi and R.~Gopakumar,
  %``Galilean Conformal Algebras and AdS/CFT,''
  arXiv:0902.1385 [hep-th].
  %%CITATION = ARXIV:0902.1385;%%


%\cite{Martelli:2009uc}
\bibitem{Martelli:2009uc}
   D.~Martelli and Y.~Tachikawa,
  %``Comments on Galilean conformal field theories and their geometric
  %realization,''
  arXiv:0903.5184 [hep-th].
  %%CITATION = ARXIV:0903.5184;%%





\end{thebibliography}
\end{document}